# Modification of special relativity and formulation of convergent and invariant quantum field theory

Jian-Miin Liu
Department of Physics, Nanjing University
Nanjing, The People's Republic of China
On leave. E-mail address: liu@mail.davis.uri.edu

Besides the two fundamental postulates, (i) the principle of relativity and (ii) the constancy of the one-way velocity of light in all inertial frames of reference, the special theory of relativity employs another assumption. This other assumption concerns the Euclidean structure of gravity-free space and the homogeneity of gravity-free time in the usual inertial coordinate system. We introduce the primed inertial coordinate system, in addition to the usual inertial coordinate system, for each inertial frame of reference, and assume the flat structures of gravity-free space and time in the primed inertial coordinate system and their generalized Finslerian structures in the usual inertial coordinate system. Combining this alternative assumption with two postulates (i) and (ii), we modify the special theory of relativity. The modified theory involves two versions of the light speed, infinite speed $\eta$ in the primed inertial coordinate system and finite speed c in the usual inertial coordinate system. It also involves the $\eta$-type Galilean transformation between any two primed inertial coordinate systems and the localized Lorentz transformation between two corresponding usual inertial coordinate systems. After confirming that all our experimental data are collected and expressed in the usual inertial coordinate system, we have the physical principle in the modified theory: the $\eta$-type Galilean invariance in the primed inertial coordinate system plus the transformation from the primed to the usual inertial coordinate systems. This principle is applied to reform of mechanics, field theory and quantum field theory. The validity of relativistic mechanics in the usual inertial coordinate system remains, while field theory is freshened. Based on the establishment of a transformation law for the quantized field systems as they are transformed from the primed to the usual inertial coordinate systems, we construct a convergent and invariant quantum field theory, in full agreement with experimental facts, founded on the modified special relativity theory and the quantum mechanics theory.

I. INTRODUCTION

The current Lorentz-invariant field theory, whether classical or quantum, has been suffering from the divergence difficulties for a long time. Indeed, the infinite self-energy of an electron in quantum electrodynamics was known as early as 1929 [1]. The origins of these difficulties lay deep within the conceptual foundations of the theory. Two foundation stones of the current quantum field theory are the special relativity theory and the quantum mechanics theory. Since it is the case that both classical and quantum field theory are plagued by the divergence difficulties, the searching direction to get to the roots of the divergence difficulties seems to be in the special relativity theory.

In 1905, Einstein published his theory of special relativity [2]. He derived the Lorentz transformation between any two usual inertial coordinate systems, which is a kinematical background for the physical principle of the Lorentz invariance. The two fundamental postulates stated by Einstein as the basis for his theory are (i) the principle of relativity and (ii) the constancy of the one-way velocity of light in all inertial frames of reference.

Besides these two fundamental postulates, the special theory of relativity also uses another assumption. This other assumption concerns the Euclidean structure of gravity-free space and the homogeneity of gravity-free time in the usual inertial coordinate system $\{x^r, t\}$, r=1,2,3, $x^1$=x, $x^2$=y, $x^3$=z:

$$dX^2 = \delta_{rs} dx^r dx^s, \quad r,s=1,2,3, \tag{1a}$$
$$dT^2 = dt^2, \tag{1b}$$

everywhere and every time.

Postulates (i) and (ii) together with the assumption Eqs.(1) yield the Lorentz transformation between any two usual inertial coordinate systems [2,3]. Indeed though the assumption Eqs.(1) was not explicitly articulated, evidently having been considered self-evident, Einstein [2] said in 1907: "Since the propagation velocity of light in empty space is c with respect to both reference systems, the two equations, $x_1^2+y_1^2+z_1^2-c^2t_1^2=0$ and $x_2^2+y_2^2+z_2^2-c^2t_2^2=0$ must be equivalent." Leaving aside a discussion of whether postulate (i) implies the linearity of transformation between any two usual inertial coordinate systems and the reciprocity of relative velocities between any two usual inertial coordinate systems, we know that the two equivalent equations, the linearity of transformation and the reciprocity of relative velocities exactly lead to the Lorentz transformation.

Some physicists explicitly articulated the assumption Eqs.(1) in their works on the topic. Pauli wrote: "This also implies the validity of Euclidean geometry and the homogeneous nature of space and time." [3]. Fock said: "The logical foundation of these methods is, in principle, the hypothesis that Euclidean geometry is applicable to real physical space together with further assumptions, viz. that rigid bodies exist and that light travels in straight lines." [4].

Introducing the four-dimensional usual inertial coordinate system $\{x^k\}$, k=1,2,3,4, $x^4$=ict, and the Minkowskian structure of gravity-free spacetime in $\{x^k\}$,

$$d\Sigma^2=\delta_{ij}dx^i dx^j, \quad i,j=1,2,3,4, \qquad (2)$$

Minkowski [5] showed in 1909 that the Lorentz transformation is just a rotation in this spacetime. He also showed how to use the four-dimensional tensor analysis for writing invariant physical laws under the Lorentz transformation. The Minkowskian structure Eq.(2) is a four-dimensional version of the assumption Eqs.(1).

In this paper, we are concerned with a modification of the special relativity theory. The modified theory keeps the two fundamental postulates (i) and (ii) but not the assumption embodied in Eqs.(1). The generalized Finslerian structures of gravity-free space and time in the usual inertial coordinate system will be assumed. Since Finsler geometry and its generalization are not familiar to many people, we begin with an introduction for them in Section II. In Section III, we learn of the definition of the usual inertial coordinate system made by Einstein and introduce the so-called primed inertial coordinate system, in addition to the usual inertial coordinate system, for each inertial frame of reference. In this section, we also make a new assumption about the local structures of gravity-free space and time. Section IV is devoted to modifying the special relativity theory. We combine this alternative assumption, instead of the assumption Eqs.(1), with the two fundamental postulates (i) and (ii). The modified theory involves two versions of the light speed, infinite speed η in the primed inertial coordinate system and finite speed c in the usual inertial coordinate system. It also involves the η-type Galilean transformation between any two primed inertial coordinate systems and the localized Lorentz transformation between two corresponding usual inertial coordinate systems. Section V addresses the four-dimensional form of the local structures of gravity-free space and time. Then, in Section VI, we turn to the physical principle in the modified theory. It is: the η-type Galilean invariance in the primed inertial coordinate system plus the transformation from the primed inertial coordinate system to the usual inertial coordinate system. The transformation from the primed to the usual inertial coordinate systems is an important part of the new physical principle. It is the characteristic of the transformation that it links the Galilean transformation between any two primed inertial coordinate systems and the localized Lorentz transformation between two corresponding usual inertial coordinate systems. For use of this physical principle, we need the knowledge about the Cartan connection in the generalized Finsler geometry. We arrange it in Section VII. Proceeding, we devote the next three sections to applications of the new physical principle to mechanics, field theory and quantum field theory. The validity of relativistic mechanics in the usual inertial coordinate system remains, while field theory and quantum field theory are freshened. In particular, a transformation law for the quantized field systems as they are transformed from the primed to the usual inertial coordinate systems is established in Section X. In Section XI, we topically discuss how the modified theory, together with the quantum mechanics theory, features a convergent and invariant quantum field theory which is in full agreement with experimental facts. Finally, we offer some concluding remarks in Section XII.

II. GENERALIZED FINSLER GEOMETRY

Finsler geometry is a kind of generalization of Riemann geometry [6,7]. It was first suggested by Riemann as early as 1854, and studied systematically by Finsler in 1918. Since then, the most significant work from the viewpoint of modern differential geometry has been done by Cartan, Rund and others.

In Finsler geometry, the distance ds between two neighboring points $x^k$ and $x^k+dx^k$, k=1,2,---,n is defined by a scale function
$$ds=F(x^1,x^2,---,x^n;dx^1,dx^2,---dx^n)$$
or simply
$$ds=F(x^k,dx^k) \tag{3}$$
which depends on directional variables $dx^k$ as well as coordinate variables $x^k$. Apart from several routine conditions like smoothness, the main constraint imposed on this scale function is that it is positively homogeneous of degree one in $dx^k$,
$$F(x^k,\lambda dx^k)=\lambda F(x^k,dx^k) \text{ for } \lambda>0. \tag{4}$$
Introducing a set of equations
$$g_{ij}(x^k,dx^k)=\partial^2 F^2(x^k,dx^k)/2\partial dx^i \partial dx^j, \ i,j=1,2,---,n, \tag{5}$$
we can represent Finsler geometry in terms of
$$ds^2=g_{ij}(x^k,dx^k)dx^i dx^j, \tag{6}$$
where $g_{ij}(x^k,dx^k)$ is called the Finslerian metric tensor. The Finslerian metric tensor is symmetric in its subscripts and all its components are positively homogeneous of degree zero in $dx^k$,
$$g_{ij}(x^k,dx^k)=g_{ji}(x^k,dx^k), \tag{7a}$$
$$g_{ij}(x^k,\lambda dx^k)=g_{ij}(x^k,dx^k) \text{ for } \lambda>0. \tag{7b}$$
We can define the so-called generalized Finsler geometry by omitting Eq.(3) as a definition of ds and instead taking Eq.(6) as a definition of $ds^2$, where the given metric tensor $g_{ij}(x^k,dx^k)$ satisfies Eqs.(7) and other routine conditions.

The generalized Finsler geometry is so-named because a Finsler geometry must be a generalized Finsler geometry but the inverse statement is not valid, in other words, a generalized Finsler geometry is not necessarily a Finsler geometry [7].

A generalized Finsler geometry with a given metric tensor $g_{ij}(x^k,dx^k)$ is also a Finsler geometry when and only when we can find a scale function $F(x^k,dx^k)$ such that this function is not only positively homogeneous of degree one in $dx^k$ and also related to $g_{ij}(x^k,dx^k)$ by Eq.(5). In this case, $F^2(x^k,dx^k)$ is positively homogeneous of degree two in $dx^k$. The following equation thus holds due to Euler's theorem on homogeneous functions,
$$2F^2(x^k,dx^k)=dx^i[\partial F^2(x^k,dx^k)/\partial dx^i]. \tag{8}$$
Iterating Eq.(8), we find
$$4F^2(x^k,dx^k)=dx^j \delta_{ij}[\partial F^2(x^k,dx^k)/\partial dx^i]$$
$$+dx^j dx^i[\partial^2 F^2(x^k,dx^k)/\partial dx^j \partial dx^i]. \tag{9}$$
It then follows from use of Eq.(8) that
$$F^2(x^k,dx^k)=dx^j dx^i[\partial^2 F^2(x^k,dx^k)/2\partial dx^j \partial dx^i]. \tag{10}$$
Eq.(10) combines with Eqs.(5) and (6) to yield Eq.(3).

III. AN ASSUMPTION ON LOCAL STRUCTURES OF SPACE AND TIME

Conceptually, the principle of relativity means that there exists a class of equivalent inertial frames of reference, any one of which moves with a non-zero constant velocity relative to any other. Einstein wrote in his Autobiographical Notes [8]: "in a given inertial frame of reference the coordinates mean the results of certain measurements with rigid (motionless) rods, a clock at rest relative to the inertial frame of reference defines a local time, and the local time at all points of space, indicated by synchronized clocks and taken together, give the time of this inertial frame of reference." As defined by Einstein, each of the equivalent inertial frames of reference is supplied with motionless, rigid unit rods of equal length and motionless, synchronized clocks of equal running rate. Then in each inertial frame of reference, an observer can employ his own motionless-rigid rods and motionless-synchronized clocks in the so-called "motionless-rigid rod and motionless-synchronized clock" measurement method to measure space and time intervals. By using this "motionless-rigid rod and motionless-synchronized clock" measurement method, the observer in each inertial frame of reference can set up his own usual inertial coordinate system $\{x^r,t\}$, r=1,2,3. Postulate (ii) means that the measured speed of light is the same constant c in every such usual inertial coordinate system. Recent null experiments searching for an anisotropy in the one-way velocity of light support postulate (ii) [9-11].

Einstein exquisitely defined the usual inertial coordinate system $\{x^r, t\}$, r=1,2,3, through associating it with the "motionless-rigid rod and motionless-synchronized clock" measurement method used in inertial frame of reference. Following Einstein, we imagine for an inertial frame of reference other measurement methods that are different from the "motionless-rigid rod and motionless-synchronized clock" measurement method. According to Einstein, by taking these other measurement methods the inertial frame of reference can set up other inertial coordinate systems, just as well as this inertial frame of reference can set up the usual inertial coordinate system by taking the "motionless-rigid rod and motionless-synchronized clock" measurement method. We call these other inertial coordinate systems the unusual inertial coordinate systems. It is understood that the relationship between the usual inertial coordinate system and any unusual inertial coordinate system of a given inertial frame of reference is specified by the relationship between the "motionless-rigid rod and motionless-synchronized clock" measurement method and the alternative measurement method associated with the mentioned unusual inertial coordinate system.

Conventional believe in flatness of gravity-free space and time is natural. But question is, in which inertial coordinate system the gravity-free space and time directly display their flatness. The special theory of relativity recognizes the usual inertial coordinate system, as shown in the assumption Eqs.(1). Making a difference, we chose one of the unusual inertial coordinate systems, say the primed inertial coordinate system $\{x'^r, t'\}$, r=1,2,3.

We assume that gravity-free space and time possess the flat metric structures in the primed inertial coordinate system, and hence, the following non-flat structures in the usual inertial coordinate system,

$$dX^2 = \delta_{rs} dx'^r dx'^s = g_{rs}(y) dx^r dx^s, \quad r,s=1,2,3, \tag{11a}$$

$$dT^2 = dt'^2 = g(y) dt^2, \tag{11b}$$

$$g_{rs}(y) = K^2(y) \delta_{rs}, \tag{12}$$

$$g(y) = (1 - y^2/c^2), \tag{13}$$

$$K(y) = \frac{c}{2y} (1 - y^2/c^2)^{1/2} \ln \frac{c+y}{c-y}, \tag{14}$$

where $y = (y^s y^s)^{1/2}$, $y^s = dx^s/dt$, s=1,2,3.

Two non-flat metric tensors $g_{rs}(y)$ and $g(y)$ depend on, actually only on, directional variables $y^s$, s=1,2,3. We call local structures specified by metric tensors $g_{rs}(y)$ and $g(y)$ the generalized Finslerian structures [12]. The generalized Finslerian structures of gravity-free space and time in the usual inertial coordinate system become flat when and only when y approaches zero.

IV. η-TYPE GALILEAN AND LOCALIZED LORENTZ TRANSFORMATIONS

We now can form a new theory by combining the assumption Eqs.(11), instead of the assumption Eqs.(1), with the two fundamental postulates (i) and (ii).

If we define a new type of velocity, $y'^s = dx'^s/dt'$, s=1,2,3, in the primed inertial coordinate system and keep the well-defined usual (Newtonian) velocity $y^s = dx^s/dt$, s=1,2,3, in the usual inertial coordinate system, we find from the assumption Eqs.(11),

$$\delta_{rs} y'^r y'^s = [\frac{c}{2y} \ln \frac{c+y}{c-y}]^2 \delta_{rs} y^r y^s, \quad r,s=1,2,3, \tag{15}$$

$$y'^s = [\frac{c}{2y} \ln \frac{c+y}{c-y}] y^s, \quad s=1,2,3, \tag{16}$$

and

$$y' = \frac{c}{2} \ln \frac{c+y}{c-y}, \tag{17}$$

where $y' = (y'^s y'^s)^{1/2}$, $y = (y^s y^s)^{1/2}$, s=1,2,3.

Two different measurement methods can be applied to a motion when it is observed in an inertial frame of reference, one being the "motionless-rigid rod and motionless-synchronized clock" measurement method, the other one being associated with the primed inertial coordinate system. As a result, two different velocities, primed velocity $y'^s$ (of the new type) and usual velocity $y^s$, are obtained. They are two versions of the motion obtained via two different measurement methods taken in the inertial frame of reference.

These two velocities are related by Eqs.(16) and (17). Velocity $y'^s$, s=1,2,3, varies uniquely with $y^s$ and equals $y^s$ only when $y^s$ vanishes. The Galilean composition among primed velocities links up with the Einstein composition among usual velocities [13]. This statement can be easily justified in the one-dimensional case:

$y'_2 = y'_1 - u' = (c/2)\ln[(c+y_1)/(c-y_1)] - (c/2)\ln[(c+u)/(c-u)]$

and

$y'_2 = (c/2)\ln[(c+y_2)/(c-y_2)]$

imply

$y_2 = (y_1-u)/(1-y_1u/c^2)$.

In Eq.(17), as y goes to c, we get an infinite speed of the new type,

$$\eta = \lim_{y \to c} \frac{c}{2} \ln \frac{c+y}{c-y} . \tag{18}$$

Speed $\eta$ is a new version of the light speed, its version in the primed inertial coordinate systems. Speed $\eta$ is invariant in the primed inertial coordinate systems owing to the invariance of speed c in the usual inertial coordinate systems.

Let IFR1 and IFR2 be two inertial frames of reference, where IFR2 moves with a non-zero constant velocity relative to IFR1. IFR1 and IFR2 can, respectively, use their own "motionless-rigid rod and motionless-synchronized clock" measurement methods and set up their own usual inertial coordinate systems $\{x^r_m, t_m\}$, m=1,2, r=1,2,3. IFR1 and IFR2 can also, respectively, set up their own primed inertial coordinate systems $\{x'^r_m, t'_m\}$, m=1,2, r=1,2,3.

Since the propagation velocity of light is $\eta$ in both primed inertial coordinate systems $\{x'^r_1, t'_1\}$ and $\{x'^r_2, t'_2\}$, we have two equivalent equations

$\delta_{rs} dx'^r_1 dx'^s_1 - \eta^2 (dt'_1)^2 = 0$, (19a)

$\delta_{rs} dx'^r_2 dx'^s_2 - \eta^2 (dt'_2)^2 = 0$. (19b)

Using Eqs.(11) with y=c, we have further two equivalent equations,

$K^2(c)\delta_{rs} dx^r_1 dx^s_1 - \eta^2 g(c)(dt_1)^2 = 0$,

$K^2(c)\delta_{rs} dx^r_2 dx^s_2 - \eta^2 g(c)(dt_2)^2 = 0$,

or

$\delta_{rs} dx^r_1 dx^s_1 - c^2(dt_1)^2 = 0$, (20a)

$\delta_{rs} dx^r_2 dx^s_2 - c^2(dt_2)^2 = 0$, (20b)

because

$c^2 K^2(c) = \eta^2 g(c)$, (21)

where

$K(c) = \lim_{y \to c} K(y)$, $g(c) = \lim_{y \to c} g(y)$.

Two equivalent equations (19), the linearity of transformation between two $\{x'^r_m, t'_m\}$, m=1,2, the reciprocity of relative primed velocities between two $\{x'^r_m, t'_m\}$, m=1,2, and the flat structures of gravity-free space and time in two $\{x'^r_m, t'_m\}$, m=1,2, will lead to the $\eta$-type Galilean transformation between two primed inertial coordinate systems $\{x'^r_m, t'_m\}$, m=1,2, under which speed $\eta$ is invariant. Two equivalent equations (20), the linearity of transformation between two $\{x^r_m, t_m\}$, m=1,2, and the reciprocity of relative usual velocities between two $\{x^r_m, t_m\}$, m=1,2, will lead to the localized Lorentz transformation between two usual inertial coordinate systems $\{x^r_m, t_m\}$, m=1,2, where the space and time differentials respectively take places of the space and time variables in the Lorentz transformation.

In the new theory, the $\eta$-type Galilean transformation stands between any two primed inertial coordinate systems, while the localized Lorentz transformation between two corresponding usual inertial coordinate systems. Substituting the assumption Eqs.(11) for the assumption Eqs.(1) does not spoil the localized Lorentz transformation between any two usual inertial coordinate systems.

V. THE FOUR-DIMENSIONAL GENERALIZED FINSLERIAN STRUCTURE

The generalized Finslerian structures Eqs.(11) can be rewritten in the four-dimensional form. To do it, following Minkowski, we introduce the four-dimensional primed inertial coordinate system $\{x'^k\}$, k=1,2,3,4, $x'^4 = i\eta t'$, and the flat structure of gravity-free spacetime in this coordinate system,

$d\Sigma^2 = \delta_{ij} dx'^i dx'^j$, i,j=1,2,3,4. (22)

Using Eqs.(11), we can find the generalized Finslerian structure of gravity-free spacetime in the four-dimensional usual inertial coordinate system $\{x^k\}$, k=1,2,3,4, $x^4$=ict,

$$d\Sigma^2 = g_{ij}(y)dx^i dx^j, \quad i,j=1,2,3,4, \tag{23a}$$

with diagonal metric tensor $g_{ij}(y)$,

$$g_{11}(y)=g_{22}(y)=g_{33}(y)=K^2(y), \quad g_{44}(y)=G^2(y), \tag{23b}$$
$$g_{ij}(y)=0 \text{ for } i \neq j, \tag{23c}$$

where

$$G(y) = \frac{\eta}{c}(1-y^2/c^2)^{1/2} \tag{24}$$

and $y=(y^s y^s)^{1/2}$, $y^s = dx^s/dt$, s=1,2,3.

It is good to represent this generalized Finslerian structure in terms of four-dimensional vector variables $z^k$, k=1,2,3,4,

$$z^r \equiv dx^r/dt' = y^r/(1-y^2/c^2)^{1/2}, \quad r=1,2,3, \tag{25a}$$
$$z^4 \equiv dx^4/dt' = ic/(1-y^2/c^2)^{1/2}. \tag{25b}$$

We have

$$d\Sigma^2 = g_{ij}(z)dx^i dx^j, \quad i,j=1,2,3,4, \tag{26a}$$

with

$$g_{11}(z)=g_{22}(z)=g_{33}(z)=K^2(z), \quad g_{44}(z)=G^2(z), \tag{26b}$$
$$g_{ij}(z)=0 \text{ for } i \neq j, \tag{26c}$$

where

$$K(z) = \frac{c}{2z} \ln \frac{\sqrt{c^2+z^2}+z}{\sqrt{c^2+z^2}-z}, \tag{27}$$

$$G(z) = \frac{c}{2\sqrt{c^2+z^2}} \lim_{z \to +\infty} \ln \frac{\sqrt{c^2+z^2}+z}{\sqrt{c^2+z^2}-z}, \tag{28}$$

and $z=(z^s z^s)^{1/2}$, s=1,2,3. In writing Eqs.(26), (27) and (28), we used two relations,

$$z/z^4 = y/ic, \tag{29}$$
$$z^4 = i(c^2+z^2)^{1/2} \text{ or } (1-y^2/c^2)^{1/2} = (1+z^2/c^2)^{-1/2}. \tag{30}$$

Metric tensor $g_{ij}(z)$ is still symmetric in its subscripts and positively homogenous of degree zero in its arguments $z^k$.

VI. THE PHYSICAL PRINCIPLE

Experiments clearly support the existence of the constancy of the usual light speed c, the Einstein addition law of relative usual velocities, and the Lorentz or the local Lorentz invariance [9-11]. These facts verify not only postulate (ii) but also that it is the "motionless-rigid rod and motionless-synchronized clock" measurement method which we use in our experiments. In other words, all our experimental data are collected and expressed in the usual inertial coordinate system.

The physical principle in the special theory of relativity is the Lorentz invariance in the usual inertial coordinate system: All physical laws keep their forms with respect to the Lorentz transformation. In the modified theory, we may also chose the usual inertial coordinate system for theoretical purposes. In doing so, the physical principle is the local Lorentz invariance: In the usual inertial coordinate system, all physical laws keep their forms with respect to the localized Lorentz transformation. But, this local Lorentz invariance must be implemented in dealing with the generalized Finslerian structures Eqs.(11).

The modified theory, however, offers us another choice for theoretical purposes. That is the primed inertial coordinate system. Making this choice, we have the physical principle: The η-type Galilean invariance in the primed inertial coordinate system plus the transformation from the primed inertial coordinate system to the usual inertial coordinate system. In other words, all physical laws are written in the η-type Galilean-invariant form in the primed inertial coordinate system; all calculations are done in the η-type Galilean-invariant manner in the primed inertial coordinate system; and all results of these calculations are finally transformed from the primed inertial coordinate system into the usual inertial coordinate system and compared to experimental facts there.

The transformation from the primed inertial coordinate system $\{x'^r, t'\}$, $r=1,2,3$, to the usual inertial coordinate system $\{x^r, t\}$, $r=1,2,3$, is an important part of the new physical principle. If two axes $x'^r$ and $x^r$, where r equals 1,2,3, are set to have the same direction and the same origin, the transformation from the primed to the usual inertial coordinate systems has the form of

$$dx'^r = K(y)dx^r, \quad r=1,2,3, \tag{31a}$$
$$dt' = (1-y^2/c^2)^{1/2} dt \tag{31b}$$

or

$$dx'^r = K(z)dx^r, \quad r=1,2,3, \tag{32a}$$
$$dt' = dt/[1+z^2/c^2]^{1/2}, \tag{32b}$$

where $K(y)$ is in Eq.(14) and $K(z)$ is in Eq.(27). Eqs.(31b) and (32b) can be rewritten respectively as

$$dx'^4 = G(y)dx^4, \tag{33}$$

and

$$dx'^4 = G(z)dx^4, \tag{34}$$

where

$$x'^4 = i\eta t' \text{ and } x^4 = ict, \tag{35}$$

and $G(y)$ is in Eq.(24) and $G(z)$ is in Eq.(28).

The transformation from the primed to the usual inertial coordinate systems contains local velocity variables $y^r$ or $z^r$, $r=1,2,3$, as parameters. It becomes identical when and only when $y=(y^r y^r)^{1/2}$ equals zero or $z=(z^r z^r)^{1/2}$ equals zero. When y approaches c or z, $z=y/(1-y^2/c^2)^{1/2}$, goes to infinity, the transformation becomes strange because of

$$K(c)=G(c)=\lim_{y \to c} \frac{c}{2y}(1-y^2/c^2)^{1/2} \ln\frac{c+y}{c-y} = 0. \tag{36}$$

Under the transformation, any contravariant vector transforms like

$$P'^r = K(z)P^r, \quad r=1,2,3, \tag{37a}$$
$$P'^4 = G(z)P^4, \tag{37b}$$

while covariant vector transforms like

$$P'_r = K^{-1}(z)P_r, \quad r=1,2,3, \tag{38a}$$
$$P'_4 = G^{-1}(z)P_4. \tag{38b}$$

It is the characteristic of the transformation Eqs.(31) that it links the Galilean transformation between any two primed inertial coordinate systems and the localized Lorentz transformation between two corresponding usual inertial coordinate systems. To justify this statement, we consider the case of one-dimensional space for simplification.

Let IFR1 and IFR2 be two inertial frames of reference, and $\{x'_m, t'_m\}$ and $\{x_m, t_m\}$ be respectively the primed inertial coordinate system and the usual inertial coordinate system of IFRm, m=1,2. Suppose IFR2 moves relative to IFR1 with an arbitrarily constant usual velocity u as measured by IFR1, and suppose an object moves relative to IFR1 with an arbitrarily usual velocity $y_1$ as measured by IFR1, and hence, relative to IFR2 with a certain specified usual velocity $y_2$ as measured by IFR2. Three primed velocities are u' and $y'_m$, m=1,2, respectively corresponding to usual velocities u and $y_m$, m=1,2.

First of all, from the transformation Eqs.(31), we know that three usual velocities u and $y_m$, m=1,2, are respectively connected to their corresponding primed velocities by

$$y'_m = (c/2)\ln[(c+y_m)/(c-y_m)], \quad m=1,2, \tag{39}$$
$$u' = (c/2)\ln[(c+u)/(c-u)]. \tag{40}$$

The Galilean transformation between $\{x'_1, t'_1\}$ and $\{x'_2, t'_2\}$ reads

$$x'_2 = x'_1 - u't'_1, \quad t'_2 = t'_1, \tag{41}$$

or localizedly

$$dx'_2 = dx'_1 - u'dt'_1, \tag{42a}$$
$$dt'_2 = dt'_1. \tag{42b}$$

We divide Eq.(42a) by Eq.(42b) for

$$y'_2 = y'_1 - u', \tag{43}$$

and insert Eqs.(39) and (40) in Eq.(43) for

$$y_2 = (y_1 - u)/[1 - y_1 u/c^2]. \tag{44}$$

Eq.(44) implies

$$[1-y_2^2/c^2]^{1/2} = [1-y_1^2/c^2]^{1/2}[1-u^2/c^2]^{1/2}/[1-y_1 u/c^2]. \tag{45}$$

Consecutively using Eqs.(31b) of m=1,2 and (45) in Eq.(42b), we can obtain

$$dt_2=\gamma(dt_1-udx_1/c^2), \tag{46a}$$

where $\gamma=1/[1-u^2/c^2]^{1/2}$ and $dx_1=y_1dt_1$. As we immediately put Eqs.(44) and (46a) into the equality $dx_2=y_2dt_2$, we can obtain another equation,

$$dx_2=\gamma(dx_1-udt_1). \tag{46b}$$

Eqs.(46a) and (46b) have the same shape as the Lorentz transformation apart from that the space and time differentials respectively take places of the space and time variables. The velocity $y_1$ can be arbitrary, so do $dx_1$ and $dt_1$. Eqs.(46) are full of meanings as the localized Lorentz transformation between $\{x_1,t_1\}$ and $\{x_2,t_2\}$.

## VII. THE CARTAN CONNECTION

Like Finsler geometry, the generalized Finsler geometry can be endowed with the Cartan connection [6,7]. Let us consider a generalized Finsler geometry with a given metric tensor $g_{ij}(x^k,dx^k)$ or $g_{ij}(x^k,z^k)$, where $z^k=dx^k/db$, $k=1,2,\cdots,n$, and $db$ is a positive invariant.

According to Cartan, the absolute differentials of vectors $X^i$ and $X_i$ can be written as

$$DX^i = dX^i + C^i_{kh}(x,z)X^k dz^h + \Gamma^i_{kh}(x,z)X^k dx^h, \tag{47a}$$
$$DX_i = dX_i - C^k_{ih}(x,z)X_k dz^h - \Gamma^k_{ih}(x,z)X_k dx^h, \tag{47b}$$

and two kinds of covariant partial derivative, $X^i_{|h}$, $X_{i|h}$, $X^i_{\|h}$, $X_{i\|h}$, can be defined by

$$DX^i = X^i_{|h}dx^h + X^i_{\|h}Dz^h, \tag{48a}$$
$$DX_i = X_{i|h}dx^h + X_{i\|h}Dz^h, \tag{48b}$$

where $C^i_{kh}$ and $\Gamma^i_{kh}$ are the affine coefficients, $Dz^k$ denotes the absolute differential of $z^k$, and abbreviations $x$ and $z$ are respectively taken for $x^k$ and $z^k$ in parentheses.

Following Cartan, as we impose several conditions including

$$C^k_{mn}(x,z)z^m dz^n = 0 \tag{49}$$

on the generalized Finsler geometry, we can find the following expressions,

$$\nabla_h X^i \equiv X^i_{|h} = \partial_h X^i - (\partial G^k/\partial z^h)(\partial X^i/\partial z^k) + \Gamma^{*i}_{kh}X^k, \tag{50a}$$
$$\nabla_h X_i \equiv X_{i|h} = \partial_h X_i - (\partial G^k/\partial z^h)(\partial X_i/\partial z^k) - \Gamma^{*k}_{ih}X_k, \tag{50b}$$
$$X^i_{\|h} = \partial X^i/\partial z^h + C^i_{kh}X^k, \tag{51a}$$
$$X_{i\|h} = \partial X_i/\partial z^h - C^k_{ih}X_k, \tag{51b}$$

with

$$\Gamma^{*k}_{ih} = \Gamma^k_{ih} - C^k_{im}(\partial G^m/\partial z^h), \tag{52a}$$
$$C^k_{jh} = g^{km}C_{jmh}, \tag{52b}$$
$$\Gamma^k_{jh} = g^{km}\Gamma_{jmh} = g^{km}\{\gamma_{jmh} - C_{hmp}(\partial G^p/\partial z^j) + C_{jhp}(\partial G^p/\partial z^m)\}, \tag{52c}$$
$$\gamma_{jkh} = \frac{1}{2}\{\partial g_{jk}/\partial x^h + \partial g_{kh}/\partial x^j - \partial g_{hj}/\partial x^k\}, \tag{52d}$$
$$G^p = \frac{1}{2}\gamma^p_{hj}z^h z^j, \tag{52e}$$
$$\gamma^p_{hj} = g^{pk}\gamma_{hkj}, \tag{52f}$$
$$C_{jmh} = \frac{1}{2}(\partial g_{jm}/\partial z^h). \tag{52g}$$

For scalar function $Q(x,z)$, two kinds of covariant partial derivative are

$$\nabla_h Q \equiv Q_{|h} = \partial_h Q - (\partial G^k/\partial z^h)(\partial Q/\partial z^k), \tag{53}$$
$$Q_{\|h} = \partial Q/\partial z^h. \tag{54}$$

If metric tensor $g_{ij}(x^k,z^k)$ is $z^k$-independent, the above Cartan connection reduces to the Levi-Civita connection in Riemann geometry. If metric tensor $g_{ij}(x^k,z^k)$ is $x^k$-independent, the covariant partial derivative of the first kind reduces to the usual partial derivative,

$$\nabla_h = \partial_h. \tag{55}$$

The metric tensor in the generalized Finslerian structure Eqs.(26) just fits in this case, i.e. it is $x^k$-independent.

## VIII. MECHANICS

It should be reminded before we go to applications of the new physical principle to mechanics and field theory that the light speed, which owns the constancy in all inertial frames of reference according to postulate (ii), has two versions: infinite speed $\eta$ in the primed inertial coordinate system and finite speed c in the usual inertial coordinate system. We deal with this constant in such a way that its value is $\eta$ in the primed inertial coordinate system and c in the usual inertial coordinate system.

The motion equation in the primed inertial coordinate system is

$$F'^k = m_0 dz'^k/dt', \quad k=1,2,3,4, \tag{56}$$

where $m_0$ is the mass of an object,

$$z'^k = dx'^k/dt', \quad k=1,2,3,4, \tag{57}$$

is the four-dimensional primed velocity and $F'^k$ is the four-dimensional force. Performing the transformation from the primed to the usual inertial coordinate systems, we get

$$F^k = m_0 Dz^k/dt', \quad k=1,2,3,4, \tag{58}$$

in the usual inertial coordinate system, where $Dz^k$ is the absolute differential of four-dimensional velocity vector $z^k$,

$$z^k = dx^k/dt', \quad k=1,2,3,4. \tag{59}$$

Dynamic variables $z^k$, $k=1,2,3,4$, are just the directional variables in the generalized Finslerian structure Eqs.(26), so we have

$$Dz^i = dz^i + C^i_{kh}(x,z)z^k dz^h + \Gamma^i_{kh}(x,z)z^k dx^h, \quad i=1,2,3,4, \tag{60}$$

according to Eq.(47a). In Eq.(60),

$$C^i_{kh}(x,z)z^k dz^h = 0, \quad i=1,2,3,4, \tag{61}$$

because of Eq.(49) and

$$\Gamma^i_{kh}(x,z) = 0, \quad i,j,k=1,2,3,4, \tag{62}$$

because of the $x^k$-independence of metric tensor $g_{ij}(z)$ in the generalized Finslerian structure Eqs.(26). We thus have $Dz^i = dz^i$, $i=1,2,3,4$, and finally,

$$F^k = m_0 dz^k/dt', \quad k=1,2,3,4, \tag{63}$$

as the motion equation in the usual inertial coordinate system.

In the primed inertial coordinate system, four-dimensional energy-momentum vector of a moving particle with rest mass $m_0$ is

$$p'^k = m_0 z'^k = (m_0 y'^1, m_0 y'^2, m_0 y'^3, im_0\eta) = (m_0 y'^r, im_0\eta), \quad r=1,2,3. \tag{64}$$

We transform it to the usual inertial coordinate system and find

$$p'^k \rightarrow p^k = (m_0 y'^r/K(y), im_0\eta/G(y))$$
$$= (m_0 y^r/[1-y^2/c^2]^{1/2}, im_0 c/[1-y^2/c^2]^{1/2})$$
$$= m_0 z^k, \quad k=1,2,3,4, \quad r=1,2,3, \tag{65}$$

as the four-dimensional energy-momentum vector in the usual inertial coordinate system, where Eqs.(14), (16) and (24) are used.

Immediately calculating the traces of tensors $p'^i p'^j$ and $p^i p^j$ yields

$$Sp(p'^i p'^j) = \begin{array}{l} -m_0^2 \eta^2, \; y'<\eta, \\ 0, \; y'=\eta, \end{array} \tag{66}$$

$$Sp(p^i p^j) = \begin{array}{l} -m_0^2 c^2, \; y<c, \\ 0, \; y=c, \end{array} \tag{67}$$

where $p'^i$ and $p^i$, $i=1,2,3,4$, are respectively in Eqs.(64) and (65). The familiar form of Eq.(67) is

$$E^2 = c^2 \overline{p} \cdot \overline{p} + m_0^2 c^4, \tag{68}$$

where

$$\overline{p} = (p^1, p^2, p^3), \quad E = -icp^4, \tag{69}$$

and E is the energy.

The motion equation in the usual inertial coordinate system has the same form as that in the special relativity theory. The four-dimensional energy-momentum vector in the usual inertial coordinate system has also the same expression as that in the special relativity theory. Four components of this vector satisfy the same relation as that in the special relativity theory. So, the change of assumption from Eqs.(1) to Eqs.(11) has no effect on the validity of relativistic mechanics in the usual inertial coordinate system.

However, the equation about invariant $\delta_{ij}p'^i p'^j = g_{ij}(z)p^i p^j$ alters,

$$\delta_{ij}p'^i p'^j = -m_0^2 \eta^2, \; y'<\eta, \tag{70}$$

$$0, \quad y'=\eta,$$

in the primed inertial coordinate system and

$$g_{ij}(z)p^i p^j = -m_0^2 c^2, \quad y<c, \qquad (71)$$
$$0, \quad y=c,$$

in the usual inertial coordinate system. This alteration will affect the forms of field equations in the usual inertial coordinate system.

IX. FIELD THEORY

The regulation that enables us promptly to write down the η-type Galilean-invariant field equations in the primed inertial coordinate system is: Having the Lorentz-invariant field equations in the current field theory, we replace speed c with speed η and add a 'prime' to each of the field variables, coordinate variables and coordinate-differentiations.

A. Field Equations and Lagrangian Densities

The interaction-free field equations and the Lagrangian densities in the primed inertial coordinate system are

$$[\delta^{ij}\partial'_i\partial'_j - æ'^2]\varphi' = 0 \text{ for massive real scale field } \varphi', \qquad (72)$$

$$\delta^{ij}\partial'_i\partial'_j A'_k = 0, \; k=1,2,3,4, \text{ for massless vector field } A'_k, \qquad (73a)$$

$$[\delta^{ij}\partial'_i\partial'_j - æ'^2]A'_k = 0, \; k=1,2,3,4, \text{ for massive vector field } A'_k, \qquad (73b)$$

$$[\delta^{ij}\gamma_i\partial'_j - æ']\psi' = 0 \text{ and} \qquad (74a)$$

$$\overline{\psi}'[\delta^{ij}\gamma_i\partial'_j - æ'] = 0 \text{ for massive even-spinor (Dirac) field } \psi', \qquad (74b)$$

and

$$L'_1 = -\frac{1}{2}\delta^{ij}\partial'_i\varphi'\partial'_j\varphi' - \frac{1}{2}æ'^2\varphi'^2 \text{ for massive real scalar field } \varphi',$$

$$L'_{2a} = -\frac{1}{4}\delta^{im}\delta^{jn}F'_{ij}F'_{mn}, \; F'_{ij} = \partial'_i A'_j - \partial'_j A'_i, \text{ for massless vector field } A'_k,$$

$$L'_{2b} = -\frac{1}{4}\delta^{im}\delta^{jn}F'_{ij}F'_{mn} - \frac{1}{2}æ'^2\delta^{ij}A'_i A'_j, \; F'_{ij} = \partial'_i A'_j - \partial'_j A'_i, \text{ for massive vector field } A'_k,$$

$$L'_3 = -\hbar c\overline{\psi}'[\delta^{ij}\gamma_i\partial'_j + æ']\psi' \text{ for massive even-spinor field } \psi',$$

where æ'=$m_0\eta/\hbar$, and $\gamma_i$, i=1,2,3,4, are the Dirac matrices, $\{\gamma_i,\gamma_j\}=2\delta_{ij}$.

As for those in the usual inertial coordinate system, we notice that Eqs.(55) and (71) imply

$$\delta^{ab}\partial'_a\partial'_b = g^{ab}(z)\nabla_a\nabla_a = g^{ab}(z)\partial_a\partial_b, \; a,b=1,2,3,4. \qquad (75)$$

The interaction-free field equations in the usual inertial coordinate system are

$$[g^{ij}(z)\partial_i\partial_j - æ^2]\varphi = 0 \text{ for massive real scale field } \varphi, \qquad (76)$$

$$\delta^{ij}\partial_i\partial_j A_k = 0, \; k=1,2,3,4, \text{ for massless vector field } A_k, \qquad (77a)$$

$$[g^{ij}(z)\partial_i\partial_j - æ^2]A_k = 0, \; k=1,2,3,4, \text{ for massive vector field } A_k, \qquad (77b)$$

$$[g^{ij}(z)\gamma_i\partial_j - æ]\psi = 0 \text{ and} \qquad (78a)$$

$$\overline{\psi}[g^{ij}(z)\gamma_i\partial_j - æ] = 0 \text{ for massive even-spinor field } \psi, \qquad (78b)$$

where æ=$m_0 c/\hbar$, $\gamma_s = K(z)\gamma'_s$, s=1,2,3, $\gamma_4 = G(z)\gamma'_4$, and $\gamma_s$, s=1,2,3, and $\gamma_4$ form the version of the Dirac matrices in the usual inertial coordinate system, $\{\gamma_i,\gamma_j\}=2g_{ij}(z)$. Here we used K(c)=G(c) in Eq.(36) and $g^{ij}(c)=K^{-2}(c)\delta^{ij}$ to get Eq.(77a). The Lagrangian densities in the usual inertial coordinate system are

$$L_1 = -\frac{1}{2}g^{ij}(z)\partial_i\varphi\partial_j\varphi - \frac{1}{2}æ^2\varphi^2 \text{ for massive real scalar field } \varphi,$$

$$L_{2a} = -\frac{1}{4}\delta^{im}\delta^{jn}F_{ij}F_{mn}, \; F_{ij} = \partial_i A_j - \partial_j A_i, \text{ for massless vector field } A_k,$$

$$L_{2b}=-\frac{1}{4}g^{im}(z)g^{jn}(z)F_{ij}F_{mn}-\frac{1}{2}æ^2 g^{ij}(z)A_i A_j,\ F_{ij}=\partial_i A_j-\partial_j A_i,\ \text{for massive vector field } A_k,$$

$$L_3=-\hbar c \overline{\psi}\,[g^{ij}(z)\gamma_i \partial_j + æ]\psi \quad \text{for massive even-spinor field } \psi.$$

## B. The de Broglie Wave Solution

It is easy to examine that the interaction-free field equations in either primed or usual inertial coordinate systems have the de Broglie wave solution,

$$\approx \exp[\frac{i}{\hbar}p'_i x'^i] = \exp[\frac{i}{\hbar}p_i x^i],\ i=1,2,3,4. \tag{79}$$

The equal mark in Eq.(79) is due to the transformation from the primed to the usual inertial coordinate systems for interaction-free particles. Actually, when we put Eqs.(79) into Eqs.(72-74) and (76-78), respectively, we get Eqs.(70) and (71) again. The de Broglie wave characteristic of interaction-free particles has been well verified by experiments.

## C. The Canonically Conjugate Variables

The canonically conjugate variables to field variables $\varphi'_k$ (vector or spinor) are defined by

$$\pi'^k = \partial L'/\partial(\partial \varphi'_k/\partial t') \tag{80}$$

in the primed inertial coordinate system. In the usual inertial coordinate system, we have

$$\pi^k = \partial L/\partial(\partial \varphi_k/\partial t'). \tag{81}$$

The expression Eq.(81) is different from $\pi^k = \partial L/\partial(\partial \varphi_k/\partial t)$ defined in the current field theory by its differentiation with respect to t', not t. Since dt' is invariant in either primed inertial coordinate systems or usual inertial coordinate systems, the conjugate variables Eqs.(80) or (81) form a contravariant vector or spinor of the same rank as the original vector or spinor.

## D. The Noether Theorem

The Noether theorem is

$$\frac{d}{dt'}[\frac{1}{i\eta}\int f'^4 d\mathbf{x}']=0 \tag{82}$$

in the primed inertial coordinate system, while

$$\frac{d}{dt'}[\frac{1}{ic}\int f^4 d\mathbf{x}']=0 \tag{83}$$

in the usual inertial coordinate system, where $d\mathbf{x}'=dx'^1 dx'^2 dx'^3$, $f'^4$ and $f^4$ are respectively the fourth component of vectors

$$f'^i=\{L'\delta^i_j - [\partial L'/\partial(\partial \varphi'_k/\partial x'^i)][\partial \varphi'_k/\partial x'^j]\}\delta x'^j +$$
$$[\partial L'/\partial(\partial \varphi'_k/\partial x'^i)]\delta\varphi'_k,\ i,j,k=1,2,3,4, \tag{84}$$

and

$$f^i=\{Lg^i_j(z) - [\partial L/\partial(\partial \varphi_k/\partial x^i)][\partial \varphi_k/\partial x^j]\}\delta x^j +$$
$$[\partial L/\partial(\partial \varphi_k/\partial x^i)]\delta\varphi_k,\ i,j,k=1,2,3,4. \tag{85}$$

In particular, the conserved field energy-momentum is

$$P'_j=\frac{1}{i\eta}\int T'^4_j d\mathbf{x}',\ j=1,2,3,4, \tag{86}$$

in the primed inertial coordinate system and

$$P_j=\frac{1}{ic}\int T^4_j d\mathbf{x}',\ j=1,2,3,4, \tag{87}$$

in the usual inertial coordinate system, where

$$T'^4_j=L'\delta^4_j-[\partial L'/\partial(\partial \varphi'_k/\partial x'^4)][\partial \varphi'_k/\partial x'^j] \tag{88}$$

and

$$T^4_j=Lg^4_j(z)-[\partial L/\partial(\partial \varphi_k/\partial x^4)][\partial \varphi_k/\partial x^j]. \tag{89}$$

Since integration element d**x**' is invariant, the conserved field energy and momentum involved in the Noether theorem are tensor components as well as $T'^4_j$ or $T^4_j$ itself in the primed inertial coordinate system and in the usual inertial coordinate system. In the current Lorentz-invariant field theory, in the usual inertial coordinate system, this characteristic has been removed by non-invariant volume element d**x**=$dx^1 dx^2 dx^3$ appearing under integration.

E. Gauge Invariance

The gauging procedure adopted in the current field theory to make a field system locally gauge-invariant with respect to a certain gauge group is still effective in the primed inertial coordinate system. For instance, in the primed inertial coordinate system, the U(1) gauge transformation is

$$\psi'_\alpha \to \psi'_\alpha \exp[i\lambda'(x')] \text{ and } A'_h \to A'_h - \frac{1}{e}\partial'_h \lambda'(x'), \tag{90}$$

and the U(1) gauge-invariant field system consists of

$$L' = L'_2 + L'_3 + L'_{int}, \tag{91}$$

with

$$L'_{int} = ie \overline{\psi}' \delta^{ij} \gamma_i A'_j \psi', \tag{92}$$

where $\lambda'(x')$ is a real scale function. Consequently, in the usual inertial coordinate system, they are

$$\psi_\alpha \to \psi_\alpha \exp[i\lambda(x)] \text{ and } A_h \to A_h - \frac{1}{e}\partial_h \lambda(x), \tag{93}$$

where Eq.(55) is used, and

$$L = L_2 + L_3 + L_{int}, \tag{94}$$

with

$$L_{int} = ie \overline{\psi} g^{ij}(z) \gamma_i A_j \psi. \tag{95}$$

Lagrangian with gauge symmetry based on the SU(2) or other non-Abelian gauge group can be also, in parallel, considered in the primed inertial coordinate system. Its version in the usual inertial coordinate system can be obtained as we act the transformation from the primed to the usual inertial coordinate systems on it.

F. Particle Size

It might be the most remarkable feature of the modified theory that the concept of particle size has its own room. In the modified theory, particle size in the primed inertial coordinate system is an invariant quantity. There is no any problem to do invariant calculations based on this quantity. The light speed $\eta$ in the primed inertial coordinate system is infinite, so it is not hard to understand any instantaneous causal processes possibly existing inside of particles. All particles properly exhibit their size in the primed inertial coordinate system. It should be, however, emphasized that the concept of particle size in the primed inertial coordinate system is different from that we tried to but failed to introduce in the current Lorentz-invariant field theory. It is quite new in its concept. We shall discuss this concept in detail in subsequent papers.

X. QUANTUM FIELD THEORY

In the primed inertial coordinate system, any field system can be quantized by use of the canonical quantization method. Instantaneity is a covariant concept, so the equal-time commutation or anti-commutation relations are reasonable. The canonically conjugate variables are contravariant to the original field variables, so the commutation or anti-commutation relations are also of tensor equations, f.g.

$$[\varphi'_\sigma(\mathbf{x}',t'), \pi'^\rho(\mathbf{x}'+\delta\mathbf{x}',t')]_\mp$$
$$= \varphi'_\sigma(\mathbf{x}',t')\pi'^\rho(\mathbf{x}'+\delta\mathbf{x}',t') \mp \pi'^\rho(\mathbf{x}'+\delta\mathbf{x}',t')\varphi'_\sigma(\mathbf{x}',t')$$
$$= i\hbar \delta_\sigma^\rho \delta^3(\delta\mathbf{x}'). \tag{96}$$

The light speed $\eta$ is infinite, so the quantum connection is acceptable to be essentially instantaneous. The primed time differential dt' is invariant in all inertial frames of reference, so the definition of time-ordered product in the perturbation expansion of S-matrix is clear.

In the primed inertial coordinate system, the state vector equation of a quantized field system is

$$i\hbar \frac{d}{dt'}|\rangle' = H'|\rangle', \tag{97}$$

while its operator evolution satisfies

$$\frac{d}{dt'}\Omega' = \frac{i}{\hbar}[H', \Omega'], \tag{98}$$

where H' is a Hamiltonian of the system, $|\rangle'$ is a state vector and $\Omega'$ is an operator. In the primed inertial coordinate system, we have a complete set of orthogonal state vectors, $\{\exp[\frac{i}{\hbar}\mathbf{p'x'}]\}$, in accordance with

$$(2\pi)^{-3}\int \exp[\frac{i}{\hbar}(\mathbf{p'}_1\mathbf{x'} - \mathbf{p'}_2\mathbf{x'})]d\mathbf{x'} = \delta^3(\mathbf{p'}_1 - \mathbf{p'}_2), \tag{99}$$

where $\mathbf{p'}=\{p'_1,p'_2,p'_3\}$, $\mathbf{x'}=(x'^1,x'^2,x'^3)$ and $\mathbf{p'x'}=p'_1x'^1+p'_2x'^2+p'_3x'^3$.

As we perform the transformation from the primed to the usual inertial coordinate systems, tensor equation Eq.(96) becomes

$$[\varphi_\sigma(\mathbf{x},t), \pi^\rho(\mathbf{x}+\delta\mathbf{x},t)] = i\hbar \frac{1}{\sqrt{g}} \delta_\sigma^{\ \rho}\delta^3(\delta\mathbf{x}), \tag{100}$$

where

$$g = \det[g_{rs}(z)], \ r,s = 1,2,3, \tag{101a}$$
$$\delta\mathbf{x'} = \sqrt{g}\,\delta\mathbf{x}, \tag{101b}$$

and Eq.(99) becomes

$$(2\pi)^{-3}\int \exp[\frac{i}{\hbar}(\mathbf{p}_1\mathbf{x} - \mathbf{p}_2\mathbf{x})]d\mathbf{x'} = \sqrt{g}\,\delta^3(\mathbf{p}_1 - \mathbf{p}_2), \tag{102}$$

due to $p'_1x'^1=p_1x^1$, $p'_2x'^2=p_2x^2$, $p'_3x'^3=p_3x^3$, for a complete set of orthogonal state vectors, $\{\exp[\frac{i}{\hbar}\mathbf{px}]\}$, in the usual inertial coordinate system, where $\mathbf{p}=(p_1,p_2,p_3)$, $\mathbf{x}=(x^1,x^2,x^3)$, and $\mathbf{px}=p_1x^1+p_2x^2+p_3x^3$.

What can we expect for Eqs.(97) and (98)? We indeed meet a transformation problem regarding state vectors and operators of the quantized field system here.

We take a real scale field $\phi'(\mathbf{x'},t')$ as an example. Its equation is

$$[\delta^{ij}\partial'_i\partial'_j - æ'^2]\phi' = 0, \ æ' = m_0\eta/\hbar, \tag{103}$$

in the primed inertial coordinate system. This equation is connected with the Lagrangian density

$$L' = -\frac{1}{2}(\delta^{ij}\partial'_i\phi'\partial'_j\phi' + æ'^2\phi'^2). \tag{104}$$

The canonically conjugate variable to $\phi'$ is

$$\pi' = \partial L'/\partial(\partial\phi'/\partial t') = (\eta^{-2})(\partial\phi'/\partial t'). \tag{105}$$

We may quantize this field assigning the equal-time commutation relations,

$$[\phi'(\mathbf{x'},t'), \phi'(\mathbf{x'}+\delta\mathbf{x'},t')] = 0, \tag{106a}$$
$$[\pi'(\mathbf{x'},t'), \pi'(\mathbf{x'}+\delta\mathbf{x'},t')] = 0 \tag{106b}$$

and

$$[\phi'(\mathbf{x'},t'), \pi'(\mathbf{x'}+\delta\mathbf{x'},t')] = i\hbar\,\delta^3(\delta\mathbf{x'}). \tag{106c}$$

Making the de Broglie wave expansion for $\phi'$,

$$\phi'(\mathbf{x'},t') = (2\pi)^{-3/2}(\hbar\eta^2/2\omega'_{\mathbf{k'}})^{1/2}\sum_{k'}\{a'_{\mathbf{k'}}\exp(-ik'_ix'^i) + a'^+_{\mathbf{k'}}\exp(+ik'_ix'^i)\}, \ i=1,2,3,4, \tag{107}$$

where $k'_i = p'_i/\hbar$, $i=1,2,3,4$, $k'_4 = i\omega'_{\mathbf{k'}}/\eta$, form the wave vector and $\mathbf{k'}=(k'_1,k'_2,k'_3)$ is called the mode parameter, we find the field energy H' and field momentum $\mathbf{P'}$,

$$H' = \sum_{k'}\hbar\omega'_{\mathbf{k'}}(a'^+_{\mathbf{k'}}a'_{\mathbf{k'}} + \frac{1}{2}), \tag{108}$$

$$\mathbf{P'} = \sum_{k'} \hbar \mathbf{k'}(a'_{\mathbf{k'}}{}^+ a'_{\mathbf{k'}}), \tag{109}$$

where $\mathbf{P'} = (P'_1, P'_2, P'_3)$, $a'_{\mathbf{k'}}{}^+$ and $a'_{\mathbf{k'}}$ are respectively the creation and annihilation operators of particle with mode $\mathbf{k'}$. They satisfy the following commutation relations,

$$[a'_{\mathbf{k'}}{}^+, a'_{\mathbf{m'}}{}^+] = [a'_{\mathbf{k'}}, a'_{\mathbf{m'}}] = 0, \tag{110a}$$

$$[a'_{\mathbf{k'}}, a'_{\mathbf{m'}}{}^+] = \delta_{\mathbf{k'm'}}. \tag{110b}$$

If we name the particle number operator of mode $\mathbf{k'}$ for

$$N'_{\mathbf{k'}} = a'_{\mathbf{k'}}{}^+ a'_{\mathbf{k'}}, \tag{111}$$

the field energy and momentum operators can be further written as

$$H' = \sum_{k'} \hbar \omega'_{\mathbf{k'}} (N'_{\mathbf{k'}} + \frac{1}{2}), \tag{112}$$

$$\mathbf{P'} = \sum_{k'} \hbar \mathbf{k'} N'_{\mathbf{k'}}. \tag{113}$$

In the particle number representation, a basis in the Hilbert space of the quantized scale field system consists of

$$|0\rangle', \text{ the vacuum state},$$
$$|1(\mathbf{k'})\rangle' = a'^+(\mathbf{k'})|0\rangle', \text{ for any } \mathbf{k'}, \text{ one-particle states},$$
$$------, \tag{114}$$
$$|n_1(\mathbf{k'}_1)---n_r(\mathbf{k'}_r)\rangle' = (n_1!---n_r!)^{-1/2}[a'^+(\mathbf{k'}_1)]^{n_1}---[a'^+(\mathbf{k'}_r)]^{n_2}|0\rangle',$$
$$\text{for any different } \mathbf{k'}_1, ---, \mathbf{k'}_r, \text{ many-particle states}.$$

These basis states are normalized in accordance with

$$'\langle n_1(\mathbf{k'}_1)---n_r(\mathbf{k'}_r)|m_1(\mathbf{j'}_1)---m_t(\mathbf{j'}_t)\rangle'$$
$$= \delta_{rt} \sum_p \delta_{n_1 m_{p(1)}} --- \delta_{n_r m_{p(t)}} \delta_{\mathbf{k'}_1 \mathbf{j'}_{p(1)}} --- \delta_{\mathbf{k'}_r \mathbf{j'}_{p(t)}}. \tag{115}$$

where the sum is over all permutation p of integers 1, ---, t.

The transformation problem is: How does a quantized field system transform under the transformation from the primed to the usual inertial coordinates systems? If we denote the operator of this quantized field system in the primed inertial coordinate system by $\Omega'$ and the state vector by $\Phi'$, and denoting those of the transformed quantized field system in the usual inertial coordinate system by $\Omega$ and $\Phi$, how are $\Omega'$ and $\Phi'$ related to $\Omega$ and $\Phi$?

This is a transformation law which answers the problem raised: Under the transformation from the primed inertial coordinate system to the usual inertial coordinate system, the quantized field system will undergo a unitary transformation for its state vectors and operators, along with just "a change" for its wave vector and other tensors. The full meaning of "a change" will be revealed below. Before we establish this law in several steps, we need to add some more to the scale field example above.

For those fields that are quantized by using anti-commutation relations, every n or m in Eqs.(114) and (115) is equal to one because the particle number operator $N'_{\mathbf{k'}}$ has only two eigenvalues in the case: zero and one. For spin non-zero fields, we have to include in spin value s', together with the mode parameter, to characterize particles. In other words, all brackets ($\mathbf{k'}$) and ($\mathbf{j'}$) in Eqs.(114) and (115) will be (s', $\mathbf{k'}$) and (i', $\mathbf{j'}$), where s' and i' are both spin parameter, and Eq.(115) will be

$$'\langle n_1(s'_1, \mathbf{k'}_1)---n_r(s'_r, \mathbf{k'}_r)|m_1(i'_1, \mathbf{j'}_1)---m_t(i'_t, \mathbf{j'}_t)\rangle'$$
$$= \delta_{rt} \sum_p \delta_{n_1 m_{p(1)}} --- \delta_{n_r m_{p(t)}} \delta_{s'_1 i'_{p(1)}} --- \delta_{s'_r i'_{p(t)}} \delta_{\mathbf{k'}_1 \mathbf{j'}_{p(1)}} --- \delta_{\mathbf{k'}_r \mathbf{j'}_{p(t)}}. \tag{116}$$

For interacting fields, the basis states Eqs.(114) will contain a number of particles of various kinds with given mode and spin, and hence, the normalization equation Eq.(116) will have more $\delta$'s on its right-hand side caused by these various kinds of particle.

Now we establish the transformation law in five steps.

Step 1. Owing to Eqs.(37a), (57) and (59), under the transformation from the primed to the usual inertial coordinate systems, the mode parameter $k'^r$, r=1,2,3, changes to $k^r$, r=1,2,3, in the way of

$$k'^r \rightarrow k^r = k'^r / K(k), \quad r=1,2,3, \tag{117}$$

where

$$K(k)=\frac{æ}{2k}\ln\{[(æ^2+k^2)^{1/2}+k]/[(æ^2+k^2)^{1/2}-k]\}, \tag{118}$$

$$k=(k^r k^r)^{1/2}, \quad r=1,2,3, \tag{119a}$$

$$k'^r=m_0 z'^r/\hbar = p'^r/\hbar, \tag{119b}$$

$$k^r=m_0 z^r/\hbar = p^r/\hbar. \tag{119c}$$

The inverse process is

$$k^r \rightarrow k'^r = K(k)k^r, \quad r=1,2,3. \tag{120}$$

This process can be clearly seen if we rewrite Eq.(120) as

$$(k,\theta,\phi) \rightarrow (f(k),\theta,\phi), \quad 0\leq\theta\leq\pi, \quad 0\leq\phi\leq 2\pi, \tag{121}$$

where

$$k^1=k\sin\theta\cos\phi, \quad k^2=k\sin\theta\sin\phi, \quad k^3=k\cos\theta,$$

and

$$k'^1=f(k)\sin\theta\cos\phi, \quad k'^2=f(k)\sin\theta\sin\phi, \quad k'^3=f(k)\cos\theta,$$

with

$$f(k)=K(k)k=\frac{æ}{2}\ln\{[(æ^2+k^2)^{1/2}+k]/[(æ^2+k^2)^{1/2}-k]\}. \tag{122}$$

This is really a contraction mapping with no change in direction. The contraction coefficient $K(k)$ depends on k itself. It becomes unity when k goes to zero and zero when k goes to infinity.

It is not hard to examine that $f(k)$ is a monotonous increasing function of k because of

$$df(k)/dk=æ/(æ^2+k^2)^{1/2} >0 \text{ for all } k. \tag{123}$$

The mapping $k^r \rightarrow k'^r$, r=1,2,3, or Eq.(121) is an one-to-one mapping with no change in direction, so do its inverse

$$k'^r \rightarrow k^r, \quad r=1,2,3,$$

and the contravariant version of this inverse,

$$k'_r \rightarrow k_r, \quad r=1,2,3, \text{ or } \mathbf{k'} \rightarrow \mathbf{k}$$

under the transformation from the primed to the usual inertial coordinate systems.

Step 2. Spin parameter is usually introduced in order to represent various types of fields, such as vector field and Dirac (even-spinor) field.

Vector field has spin components +1, 0, -1. These spin components correspond to circularly polarized field states. Spin +1 (-1) means the projection of a right (left) circular polarized field state onto the direction of motion, while spin 0 corresponds to the longitudinal field state. Dirac field has spin components +1/2 and -1/2. The quanta distinguished by spin +1/2 and spin -1/2 also differ in polarization or helicity projection onto the direction of motion or the opposite direction of motion.

It is an observation: Since three pairs of axes $x'^r$ and $x^r$, r=1,2,3, each has the same direction and there is no difference between the directions of mode parameters $k'^r$ and $k^r$, r=1,2,3, the direction of motion, and hence, the polarization projections of field states onto the direction of motion have no change under the transformation from the primed to the usual inertial coordinate systems. Spin components do not change under the transformation from the primed to the usual inertial coordinate systems at all.

Step 3. The transformation from the primed to the usual inertial coordinate systems reflects the change of measurement method from one associated with the primed inertial coordinate system to the "motionless-rigid rod and motionless-synchronized clock" measurement method. This change in measurement method does not create or annihilate any particle and does not change any particle from one kind to another kind. So, the transformation from the primed to the usual inertial coordinate systems does not create or annihilate any particle and does not change any particle from one kind to another kind, either. Because mode parameter $\mathbf{k'}$ changes to $\mathbf{k}$ in the one-to-one manner, the number of particle of a kind with mode $\mathbf{k'}$ and spin s' in the primed inertial coordinate system keeps unchanged under the transformation from the primed to the usual inertial coordinate systems, the same as that of the kind with mode $\mathbf{k}$ and spin s in the usual inertial coordinate system, where s=s' according to Step 2.

Step 4. As we perform the transformation from the primed to the usual inertial coordinate systems, since of Step 3, the basis states Eqs.(114) become

$|0\rangle$, the vacuum state,

$|1(s,\mathbf{k})\rangle$, for any $(s,\mathbf{k})$, one-particle states,

------, (124)

$|n_1(s_1,\mathbf{k}_1)\text{---}n_r(s_r,\mathbf{k}_r)\rangle$, for any different $(s_1,\mathbf{k}_1)$, ---, $(s_r,\mathbf{k}_r)$, many-particle states,

where

$s=s'$, $s_1=s'_1$, ---, $s_r=s'_r$, (125a)

$\mathbf{k}' \to \mathbf{k}$, (125b)

$\mathbf{k}'_1 \to \mathbf{k}_1$, ---, $\mathbf{k}'_r \to \mathbf{k}_r$, (125c)

and

different $(s_1,\mathbf{k}_1)$, ---, $(s_r,\mathbf{k}_r)$ come from different $(s'_1,\mathbf{k}'_1)$, ---, $(s'_r,\mathbf{k}'_r)$. (125d)

Obviously, the states Eqs.(124) are still orthogonal to each other and form a complete set. Actually, since of Eqs.(125) and Step 3, we have from Eq.(116),

$$\langle n_1(s_1,\mathbf{k}_1)\text{---}n_r(s_r,\mathbf{k}_r)|m_1(i_1,\mathbf{j}_1)\text{---}m_t(i_t,\mathbf{j}_t)\rangle$$
$$=\delta_{rt}\sum_p \delta_{n_1 m_{p(1)}} \text{---} \delta_{n_r m_{p(t)}} \delta_{s_1 i_{p(1)}} \text{---} \delta_{s_r i_{p(t)}} \delta_{k_1 j_{p(1)}} \text{---} \delta_{k_r j_{p(t)}}. \quad (126)$$

The states Eqs.(124) form another basis. The basis Eqs.(114) in the Hilbert space of the quantized field system changes to the basis Eqs.(124) under the transformation from the primed to the usual inertial coordinate systems. That means there is a unitary transformation U connecting two bases,

$|0\rangle'=U|0\rangle$, $|1(s',\mathbf{k}')\rangle'=U|1(s,\mathbf{k})\rangle$, ------,

$|n_1(s'_1,\mathbf{k}'_1)\text{---}n_r(s'_r,\mathbf{k}'_r)\rangle'=U|n_1(s_1,\mathbf{k}_1)\text{---}n_r(s_r,\mathbf{k}_r)\rangle$,

and also

$a'^+(s',\mathbf{k}')=Ua^+(s,\mathbf{k})U^{-1}$, $a'(s',\mathbf{k}')=Ua(s,\mathbf{k})U^{-1}$, $N'(s',\mathbf{k}')=UN(s,\mathbf{k})U^{-1}$.

The above argument is valid for the system of interacting fields. In the case, the unitary transformation U connects two bases of the original and transformed quantized field system, both consist of the basis states with a number of particles of various kinds. The interaction Hamiltonian is equal to the interaction Lagrangian if not counting a sign, so it can be also represented in terms of the particle creation and annihilation operators of various kinds. We thus have

$H'_{int}=UH_{int}U^{-1}$.

In general, for the quantized field system, we can find a unitary transformation U, and

$\Omega'=U\Omega U^{-1}$, (127)

$\Phi'=U\Phi$, (128)

under the transformation from the primed to the usual inertial coordinate systems.

Step 5. Acting the unitary transformation on the following eigenvalue equations,

$$\mathbf{P}'|\text{---},n_{\mathbf{k}'},\text{---}\rangle' = \sum_{k'} n\hbar \mathbf{k}'|\text{---},n_{\mathbf{k}'},\text{---}\rangle',$$

$$H'|\text{---},n_{\mathbf{k}'},\text{---}\rangle' = \sum_{k'} (n\hbar \omega'_{\mathbf{k}'} + 1/2)|\text{---},n_{\mathbf{k}'},\text{---}\rangle',$$

in the primed inertial coordinate system, we get

$$U\mathbf{P}|\text{---},n_{\mathbf{k}},\text{---}\rangle = \sum_{k} n\hbar \mathbf{k}' U|\text{---},n_{\mathbf{k}},\text{---}\rangle,$$

$$UH|\text{---},n_{\mathbf{k}},\text{---}\rangle = \sum_{k} (n\hbar \omega'_{\mathbf{k}'} + 1/2)U|\text{---},n_{\mathbf{k}},\text{---}\rangle.$$

It is necessary to change also the wave vector involved in the expansion coefficients, from $(\mathbf{k}',\omega'_{\mathbf{k}'})$ to $(\mathbf{k},\omega_{\mathbf{k}})$, if we want correct eigenvalue equations in the usual inertial coordinate system, i.e.

$$\mathbf{P}|\text{---},n_{\mathbf{k}},\text{---}\rangle = \sum_{k} n\hbar \mathbf{k}|\text{---},n_{\mathbf{k}},\text{---}\rangle,$$

$$H|\text{---},n_{\mathbf{k}},\text{---}\rangle = \sum_{k} (n\hbar \omega_{\mathbf{k}} + 1/2)|\text{---},n_{\mathbf{k}},\text{---}\rangle.$$

That is all five steps. The transformation law differently treats the operators or state vectors and the wave vector or other tensors. It contains a unitary transformation for the operators or state vectors, while just "a change" for the wave vector or other tensors. Here, "a change" means substituting the versions of the mentioned vector or tensors in the usual inertial coordinate system for their versions in the primed inertial coordinate system. The wave vector changes from $(\mathbf{k'},\omega'_{\mathbf{k'}})$ to $(\mathbf{k},\omega_{\mathbf{k}})$, or equivalently, from $(k'^r,\omega')$ to $(k^r,\omega)$, r=1,2,3, $k'^4=i\omega'/\eta$, $k^4=i\omega/c$. According to Eqs.(68) and (119), components $k^r$, r=1,2,3, and $k^4=i\omega/c$ in the usual inertial coordinate system satisfy

$$\omega^2=c^2 k^r k^r + æ^2 c^2, \quad r=1,2,3, \quad æ=m_0 c/\hbar. \tag{129}$$

The transformation law allows us immediately to write, from Eqs.(97) and (98),

$$i\hbar \frac{d}{dt'}|\rangle = H|\rangle \tag{130}$$

and

$$\frac{d}{dt'}\Omega = \frac{i}{\hbar}[H, \Omega] \tag{131}$$

as the state vector equation and operator equation in the usual inertial coordinate system. It also allows us to get Eq.(100) again from Eq.(96).

The following correspondence regulation naturally flows from the transformation law:

Having a quantized field system in the current Lorentz-invariant field theory represented in the usual inertial coordinate system, we replace speed c with speed η and add a 'prime' to each of the state vectors, operators, wave vector and other tensors, and obtain the quantized field system in the primed inertial coordinate system. We do η-type Galilean-invariant calculations for the quantized field system and have theoretical results in the primed inertial coordinate system. We perform the transformation from the primed to the usual inertial coordinate systems on these results. In doing so, we expect a unitary transformation for its all state vectors and operators, and "a change" for its wave vector and other tensors, from their versions in the primed inertial coordinate system to their versions in the usual inertial coordinate system. The components of wave vector $k^i$, i=1,2,3,4, $k^4=i\omega/c$, in the usual inertial coordinate system satisfy Eq.(129). The obtained results in the usual inertial coordinate system are to be compared to experimental facts.

XI. CONVERGENT AND INVARIANT QUANTUM FIELD THEORY

The divergence difficulties in the current quantum field theory have been already ascribed to the model of point particle. But any attempt to assign a finite size to a particle makes us face other serious difficulties [14]. In the framework of the special relativity theory, a finite size of a moving particle is not a covariant concept. It has been unknown how to carry out the Lorentz-invariant calculations based on such a size. Secondly, the Lorentz invariance recognizes the light speed c constituting a limitation for transport of matter or energy and transmission of information or causal connection. The law of causality rejects any instantaneous processes between two events at distinct space points. It is very hard to explain how a faintly sized particle as a whole is set in motion when a force acts on it at its edge. It is also hard to explain how the quantum connection specified by the quantum mechanics theory can be instantaneous. Moreover, as pointed out by Poincare [15], the field energy and momentum of a faintly sized electron in the Lorentz-invariant field theory do not have correct transformation properties, though they are finite. The field energy and momentum of a faintly sized electron do not form a vector, unlike its mechanical energy and momentum. When we compute some quantity, for instance, its total mass, the field energy and momentum will result in different values.

Picturing electron as a distribution of charge in both space and time, and introducing non-local interaction in the Lorentz-invariant manner, some physicists proposed a finite and quantizable field theory [16]. This theory is questioned because it loses the gauge invariance and still has the difficulties of the vacuum polarization divergence. It is questioned also because of inconsistency with the fact that no particle with a fraction charge has been yet observed. We would have seen such particles if the charge distribution existed. What we have seen is that all fundamental particles which are charged have the same charge, apart from a sign.

The circumstance quite alters in the theory presented in the context. The particle size is now a covariant concept in the primed inertial coordinate system. There is no problem in doing the η-type Galilean-invariant calculations based on it in the primed inertial coordinate system. As the light speed η is infinite in the primed inertial coordinate system, the causality law is always obeyed, even inside of the faintly sized particles. Instantaneous quantum connection in the primed inertial coordinate system is understandable. Moreover, the field energy and momentum of a faintly sized particle form a vector in the primed inertial coordinate system, as well as its mechanical energy and momentum, so they have correct transformation properties. The difficulties pointed out by Poincare no longer exist. In the primed inertial coordinate system, all particles properly exhibit their size. As a matter of course, we can have a convergent and η-type Galilean-invariant quantum field theory in the primed inertial coordinate system.

When we transform this convergent and invariant quantum field theory from the primed to the usual inertial coordinate systems, all its state vectors and operators undergo a unitary transformation, while the wave vector and other tensors undergo "a change" to their versions in the usual inertial coordinate system. The unitary transformation does not change the eigenvalue spectrums of its operators, the expectation values of its observations and its operator and state vector equations. We shall therefore have a convergent and invariant quantum field theory in the usual inertial coordinate system. Furthermore, as we proved in Section IV, the localized Lorentz transformation stands between any two usual inertial coordinate systems, so we shall in fact have a convergent and locally Lorentz-invariant quantum field theory in the usual inertial coordinate system.

In the current Lorentz-invariant quantum field theory, replacing speed c with speed η and adding a prime to each of the state vectors, operators, wave vector and other tensors, we can have the quantum field theory in the primed inertial coordinate system. This quantum field theory is the same as the current quantum field theory in all aspects except for speed η and those primes. For this quantum field theory, we can do almost everything that we did in the current quantum field theory, including the renormalization program. If we do so, we shall have the same results for this quantum field theory in the primed inertial coordinate system as for the current quantum field theory still except for speed η and the primes. When we further transform the results for this quantum field theory from the primed to the usual inertial coordinate systems, as a result, all added primes have gone and the light speed has its value of c. We finally have exactly the same results for this quantum field theory in the usual inertial coordinate system as for the current quantum field theory.

However, we do not suggest to do so. Yes, it can not be denied that the renormalization technique has had great experimental successes against the divergence difficulties. But it can not be also denied that renormalizability is not a basic principle of physics. Especially, since "renormalization of a quantity gives up any possibility of calculating that quantity" [17], the renormalized quantum field theory is unable to explain a class of important phenomena in particle physics--- the mass difference in some particle groups such as proton and neutron, the π-mesons, the K-mesons, the Σ-hyperons, the Ξ-hyperons. The particle size in the primed inertial coordinate system is a proper concept which ought to be involved in the quantum field theory. In such quantum field theory, renormalization technique is still effective as it is applied with finite renormalization factors.

XII. CONCLUDING REMARKS

Besides the two fundamental postulates, (i) the principle of relativity and (ii) the constancy of the light speed in all inertial frames of reference, the special theory of relativity uses another assumption. This other assumption concerns the Euclidean structure of gravity-free space and the homogeneity of gravity-free time in the usual inertial coordinate system. We have introduced the so-called primed inertial coordinate system, in addition to the usual inertial coordinate system, for each inertial frame of reference. We have proposed assuming the flat structures of gravity-free space and time in the primed inertial coordinate system, and hence, their generalized Finslerian structures in the usual inertial coordinate system. Combining this alternative assumption with the two fundamental postulates (i) and (ii), we have modified the special relativity theory.

The modified theory involves two versions of the light speed, infinite speed η in the primed inertial coordinate system and finite speed c in the usual inertial coordinate system. It also involves the η-type Galilean transformation between any two primed inertial coordinate systems and the localized Lorentz transformation between two corresponding usual inertial coordinate systems. The concept of particle size in the primed inertial coordinate system has its own place in the modified theory. The physical principle is no longer the Lorentz invariance in the usual inertial coordinate system. It is: the η-type Galilean invariance in the primed inertial coordinate system plus the transformation from the primed inertial coordinate system to the usual inertial coordinate system.

We have applied this new physical principle to reform mechanics, field theory and quantized field theory. The validity of relativistic mechanics in the usual inertial coordinate system remains, while field theory is freshened. Any field system can be quantized by using the canonical quantization method in the primed inertial coordinate system. Under the transformation from the primed to the usual inertial coordinate systems, any quantized field system will undergo a unitary transformation for its state vectors and operators, and just "a change" for its wave vector and other tensors. We have constructed a convergent and invariant quantum field theory, in full agreement with experimental facts, founded on the modified theory of special relativity and the quantum mechanics theory.

ACKNOWLEDGMENT
The author greatly appreciates the teachings of Prof. Wo-Te Shen. The author thanks Prof. Mark Y. Mott and Dr. Allen E. Baumann for helpful suggestions.

REFERENCES
[1] W. Heisenberg and W. Pauli, Z. Phys., 56, 1 (1929); 59, 168 (1930)
 I. Waller, Z. Phys., 62, 673 (1930)
 J. R. Oppenheimer, Phys. Rev., 35, 461 (1930)
 W. Heisenberg, Z. Phys., 90, 209 (1934)
 P. A. M. Dirac, Pro. Camb. Phil. Soc., 30, 150 (1934)
[2] A. Einstein, Ann. Physik, 17, 891 (1905)
 A. Einstein, Jarbuch der Radioaktivitat und Elektronik, 4, 411 (1907), reprinted in The Collected Papers of A. Einstein, vol.2, 252, Princeton University Press, Princeton, NJ (1989)
[3] A. Einstein, H. A. Lorentz, H. Minkowski and H. Weyl, The Principle of Relativity, collected papers with notes by A. Sommerfield, Dover, New York (1952)
 A. S. Eddington, The Mathematical Theory of Relativity, Cambridge UniversityPress, Cambridge (1952)
 C. Moller, The Theory of Relativity, Oxford and New York (1952)
 W. Pauli, Theory of Relativity, Pergamom Press Ltd., New York (1958), English translator: G. Field
[4] V. Fock, The Theory of SpaceTime and Gravitation, Pergamon Press (New York (1959)
[5] H. Minkowski, Phys. Z., 10, 104 (1909)
[6] P. Finsler, Uber Kurven und Flachen in Allgemeinen Raumen, Dissertation, Gottingen 1918, Birkhauser Verlag, Basel (1951)
 E. Cartan, Les Espaces de Finsler, Actualites 79, Hermann, Paris (1934)
 H. Rund, The Differential Geometry of Finsler Spaces, Springer-Verlag, Berlin (1959)
[7] G. S. Asanov, Finsler Geometry, Relativity and Gauge Theories, D. Reidel Publishing Company, Dordrecht (1985)
[8] A. Einstein, Autobiographical Notes, in: A. Einstein: Philospheo-Scientist, ed. P. A. Schipp, 3rd edition, Tudor, New York (1970)
[9] R. M. Barnett et al, Rev. Mod. Phys., 68, 611(1996), p655
[10] J. D. Prestage et al, Phys. Rev. Lett., 54, 2387(1985)
 S. K. Lamoreaux et al, Phys. Rev. Lett., 57, 3125(1986)
 T. E. Chupp et al, Phys. Rev. Lett., 63, 1541(1989)
 M. D. Garbriel and M. P. Haugan, Phys. Rev. D41, 2943(1990)
[11] A. Brillet and J. L. Hall, Phys. Rev. Lett., 42, 549(1979)
 E. Riis et al, Phys. Rev. Lett., 60, 81(1988); 62, 842(1989)
 D. Hils and J. L. Hall, Phys. Rev. Lett., 64, 1697(1990)
 T. P. Krisher et al, Phys. Rev., D45, 731(1990)
 C. M. Will, Phys. Rev., D45, 403(1992)
 R. W. McGowan et al, Phys. Rev. Lett., 70, 251(1993)
[12] Jian-Miin Liu, Galilean Electrodynamics (Massachusetts, USA), 8, 43 (1997)

[12] Jian-Miin Liu, The Properness Principle of Natural Philosophy, to be published
[13] Jian-Miin Liu, On the Fock velocity-space, to be published
[14] W. Heitler, Physical aspects of quantum-field theory, in The Quantum Theory of Fields, ed. R. Stops, Interscience Publishing Co., New York (1962)
[15] H. Poincare, Comtes rendus (Paris), $\underline{40}$, 1504 (1905); La Mechanique Nouvelle, Gauthier-Villars, Paris (1924)
[16] H. McManus, Proc. Roy. Soc. (London), A$\underline{195}$, 323 (1948)
R. P. Feynman, Phys. Rev., $\underline{74}$, 1430 (1948); $\underline{76}$, 939 (1948)
[17] R. P. Feynman, The present status of quantum electrodynamics, in The Quantum Theory of Fields, ed. R. Stops, Interscience Publishing Co., New York (1962)